\documentclass{article}
\usepackage{graphicx}
\usepackage{balance}  
\usepackage{multirow}
\usepackage{hyperref}
\usepackage{color}
\usepackage[title]{appendix}
\usepackage{fullpage}

\providecommand{\keywords}[1]{\textbf{\textit{Keywords:}} #1}

\begin{document}

\title{Cut to Fit: Tailoring the Partitioning to the Computation}
\date{FORTH Technical Report: TR-469-MARCH-2018}

\author{
Iacovos Kolokasis\\\texttt{kolokasis@ics.forth.gr}
\and
Polyvios Pratikakis\\\texttt{polyvios@ics.forth.gr}
}

\maketitle

\begin{abstract}
    Social Graph Analytics applications are very often built using
    off-the-shelf analytics frameworks.  These, however, are profiled
    and optimized for the general case and have to perform for all
    kinds of graphs.  This paper investigates how knowledge of the
    application and the dataset can help optimize performance with
    minimal effort.  We concentrate on the impact of partitioning
    strategies on the performance of computations on social graphs.  We
    evaluate six graph partitioning algorithms on a set of six social
    graphs, using four standard graph algorithms by measuring a set of
    five partitioning metrics.

    We analyze the performance of each partitioning strategy with respect
    to (i) the properties of the graph dataset, (ii) each analytics
    computation,
    of partitions.  We discover that communication cost is the best
    predictor of performance for most ---but not all--- analytics
    computations.  We also find that the best partitioning strategy
    for a particular kind of algorithm may not be the best for
    another, and that optimizing for the general case of all
    algorithms may not select the optimal partitioning strategy for a
    given graph algorithm.  We conclude with insights on selecting the
    right data partitioning strategy, which has significant impact on
    the performance of large graph analytics computations; certainly
    enough to warrant optimization of the partitioning strategy to the
    computation and to the dataset.

\end{abstract}


\section{Introduction}
\label{sec:intro}

Social Network Analytics computations are a significant part of a
growing number of big data applications in many fields.  Many
different analytics frameworks target graph analytics
computations~\cite{graphx,giraph,galois,powergraph,graphlab,graphChi},
offering high-level programming models for the development of complex
graph algorithms. 

Partitioning and placement play a much more important role in graph
computations compared to map-reduce analytics~\cite{mapReduce}, as
graph algorithms very often have irregular computational dependencies
that depend on the graph structure.  Sub-optimal partitioning and
placement of the graph may cause load-imbalance, incur high
communication costs, or delay the convergence of an iterative
computation.  Thus, many graph analytics frameworks provide a way for
the user to control partitioning, aiming to allow for
algorithm-specific optimizations.  Overall, graph partitioning
strategies are either edge-cuts or vertex-cuts~\cite{rahimian}.  Edge
cuts partition the vertex set, optimizing for the number of edges that
cross partition boundaries, as these translate to communication costs.
Abou-Rjeili and Karypis~\cite{karypis} have shown edge cuts to produce
partitions of very different sizes and may lead to load imbalance
especially for power-law graphs, as are most of the social network
analytics datasets.  To avoid such imbalance, vertex cuts divide edges
into balanced partitions, replicating vertices as required.  In this
case, communication cost tends to be proportional to the number of
replicated edges, therefore vertex cut algorithms aim to minimize
them.

However, communication cost is not perfectly correlated with the
number of cut edges or replicated vertices.  Other metrics that affect
performance include the vertices of the largest partition, the ratio
of replicated vertices to total vertices, the ratio of edges in the
largest partition to total edges, and more~\cite{metrics}.  Although
these metrics correlate with performance for many analytics
computations, it is not always straightforward to predict which one is
the most important factor for every graph algorithm.

This work aims to improve understanding of the impact partitioning
strategies have on the performance of each graph algorithm and
dataset, to help recover some of the performance lost to the
generalizations graph analytics frameworks are forced to do to provide
high-level abstractions.  We do that in GraphX~\cite{graphx}, a graph
analytics framework that is part of Apache Spark~\cite{spark}, a
popular open-source analytics engine\footnote{We selected GraphX/Spark
as these have currently the most active communities in terms of
repository commits on github.com and technical discussions on
stackoverflow.com.}.  To do that, we use a set of five partitioning
metrics to measure and compare the partitioning algorithms available
in GraphX, together with two partitioning algorithms we propose, on
six large graphs.  Moreover, we use a set of four well-known graph
algorithms to evaluate the impact of partitioning on performance, and
also to understand the predictive quality of the metrics used to
compare partitioning algorithms. 

Overall, the contributions of this paper are:
\begin{itemize}
\item We systematically evaluate a set of partitioning algorithms
    using a wide set of metrics on a set of social graphs over a set
        of four very different algorithms, implemented in GraphX.  

\item We propose two new hash partitioning algorithms that optimize a
    different combination of metrics compared to the existing vertex
        cut strategies implemented in GraphX.

\item We show which partition metric is correlated according the graph
    algorithm and the dataset and 

\item We show that partitioning depends on: (i) the number of
    partitions, (ii) the application operation and (iii) the
        properties of the graph.
\end{itemize}

We believe that our conclusions will help analytics experts optimize
their analytics pipelines to the dataset and extract much of the
performance lost to high-level abstractions, without having to resort
to custom implementations.

\section{Datasets}

\begin{figure}[t]
\begin{minipage}{0.5\textwidth}
  \centering
  \includegraphics[width=\textwidth]{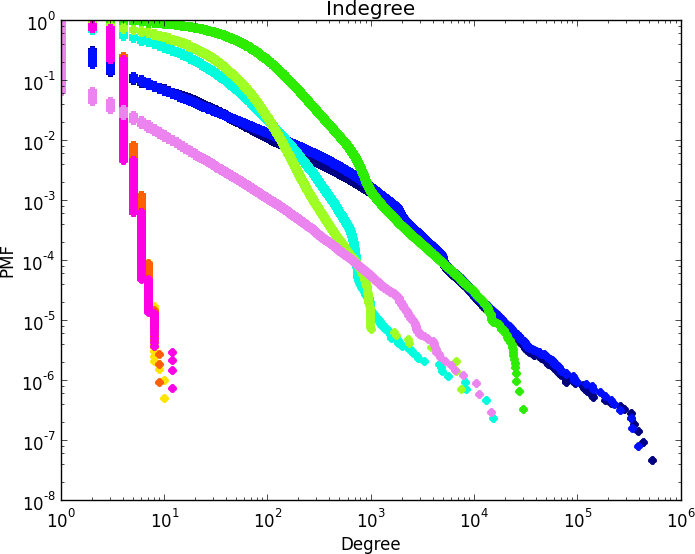}
\end{minipage}
\hfill
\begin{minipage}{0.5\textwidth}
  \centering
  \includegraphics[width=\textwidth]{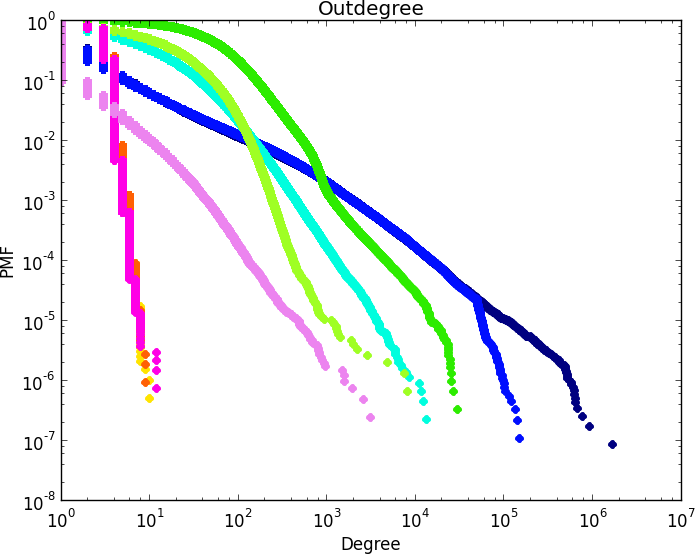}
\end{minipage}
\hfill
    \begin{minipage}{\textwidth}
        \centering
        \includegraphics[width=\textwidth]{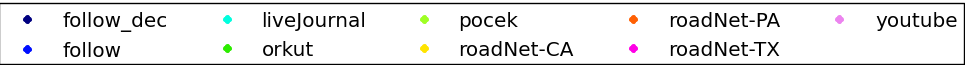}
    \end{minipage}
\caption{In-degree and Out-degree distribution of graph datasets.}
\label{fig:graph:degrees}
\end{figure}

We use a set of six social graphs to study the effect of partitioning
on social network analytics.  Four of those were obtained from the
SNAP collection of datasets referring to social networks~\cite{snap}.
Namely, %
\emph{YouTube} is a connected part of the YouTube social graph that
includes several communities~\cite{youtubedata},
\emph{Pocek} is a connected and anonymized part of the Pocek, an
on-line social network in Slovakia~\cite{pocekdata},
\emph{Orkut} is a connected part of the Orkut free on-line social
network~\cite{youtubedata},
\emph{socLiveJournal} is a sample of the livejournal.com
graph~\cite{backstrom2006group},
\emph{RoadNet-CA} is the road network of California~\cite{road},
\emph{RoadNet-PA} is the road network of
Pennsylvania~\cite{road},
\emph{RoadNet-TX} is the road network of
Texas~\cite{road}.

The \emph{follow-jul} and \emph{follow-dec} datasets are parts of the
twitter.com follow graph that we crawled using the twitter API
starting July 2016, and up to July 2017 and December 2017
respectively; and that we have anonymized by hashing user IDs.  The
first dataset is a subset of the second, and both include friend and
follower relations of any users that have published tweets in the
Greek language during the corresponding time period.

\begin{table}[t]
    \begin{center}
    \resizebox{\textwidth}{!}{
    \begin{tabular}{|c|c|c|c|c|c|c|c|c|c|}
        \hline
        Dataset        &   Vertices &  Edges &   Symm & ZeroIn\% & ZeroOut\% & Triangles & Conn.Comp. & Diameter &  Size \\ 
        \hline
        \hline
        RoadNet-PA     &       1.0M &   3.0M & 100.00 &   0.00 &    0.00 &     67.1K &    1052 & $\infty$ &83.7M \\ \hline 
        YouTube        &       1.1M &   2.9M & 100.00 &   0.00 &    0.00 &      3.0M &       1 &       20 &74.0M \\ \hline
        RoadNet-TX     &       1.3M &   3.8M & 100.00 &   0.00 &    0.00 &     82.8K &    1766 & $\infty$ &56.5M \\ \hline 
        Pocek          &       1.6M &  30.6M &  54.34 &   6.94 &   12.25 &     32.5M &       1 &       11 & 404M \\ \hline
        RoadNet-CA     &       1.9M &   5.5M & 100.00 &   0.00 &    0.00 &    120.6K &    1052 & $\infty$ &83.7M \\ \hline
        Orkut          &       3.0M & 117.1M & 100.00 &   0.00 &    0.00 &    627.5M &       1 &        9 & 3.3G \\ \hline
        socLiveJournal &       4.8M &  68.9M &  75.03 &   7.39 &   11.12 &    285.7M &   1,876 & $\infty$ & 1.0G \\ \hline
        follow-jul     &      17.1M & 136.7M &  37.57 &  46.94 &   25.65 &      4.8B &      52 & $\infty$ & 2.7G \\ \hline
        follow-dec     &      26.3M & 204.9M &  37.57 &  55.05 &   18.34 &      7.6B &      47 & $\infty$ & 4.1G \\ \hline
        
    \end{tabular}
    }
    \end{center}

    \caption{Characterization of datasets.}
    \label{tab:datasets}
\end{table}

Table~\ref{tab:datasets} shows some representative characteristics of
the datasets, as reported in the corresponding publications, or, when
missing, as we measured them using GraphX.
The first column shows the name of each dataset, ordered by the number
of vertices.
The second and third columns show the size of each graph, its number
of vertices and edges, respectively.
The fourth column shows edge symmetry, i.e., the percentage of edges
that were reciprocated.  YouTube, Facebook, Orkut and Friendster are
undirected graphs, hence their symmetry is by definition 100\%.  A
high degree of symmetry in social networks affects the structure of
the network, resulting in increased connectivity and reduced diameter
of the graph.
The fifth and sixth columns show the percentage of vertices that have
no incoming or outgoing edges, respectively.  Such ``leaf'' vertices
often occur when sampling a larger graph using forest-fire crawling to
sample the graph.
The seventh column shows the total number of triangles in each
network. The number of triangles is commonly used to assess the
density of the network.  As such, we expect datasets with higher
triangle counts to incur higher communication costs in BSP
computations.
The eighth column shows the number of connected components in each
graph.  For directed graphs, we measured connected components using
the strongly connected components algorithm implemented in GraphX.
The ninth column shows the diameter of each graph, i.e., its longest
shortest path. The diameters of the Facebook, Twitter, YouTube and
Orkut graphs are very short, as these are dense networks.  The
remaining datasets have infinite diameter, as the graphs include more
than one connected components.
The last column shows the size of the each graph dataset on disk.

\begin{figure}[t]
    \centering
    \includegraphics[width=0.7\textwidth]{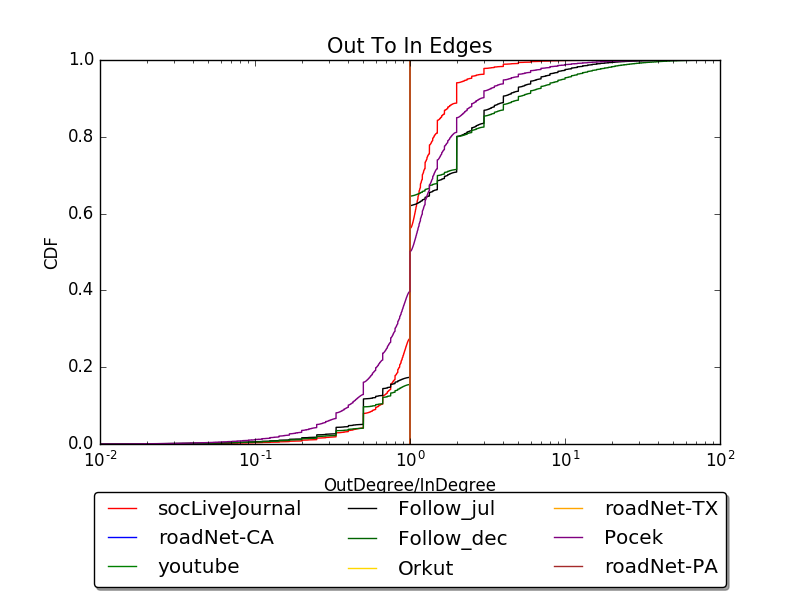}
    \caption{The CDF of the out-degree to in-degree ratio over all vertices.}
    \label{fig:outToInDegrees}
\end{figure}

To further characterize and understand each dataset,
Figure~\ref{fig:graph:degrees} shows the distribution of in-degree and
out-degree for each graph.  Although all datasets exhibit fat-tailed
distributions of both in-degree and out-degree, not all seem to be
power-law distributions.
Moreover, vertices in social graphs that have many outgoing links tend
to have many incoming links.  Not all networks, however, exhibit this
pattern to the same degree.  We compare the graphs in this regard, by
computing the distribution of the ratio of out-degree to in-degree
over all vertices.  Figure~\ref{fig:outToInDegrees} shows the
cumulative distribution function for this ratio.  Note that for
Facebook and YouTube all vertices have a ratio of 1, since they are
undirected graphs.  All directed graphs exhibit the pattern where most
users have a very similar number of in- and out-edges, although the
socLiveJournal graph has the least number of ``superstar'' users
compared to the rest.  Twitter and Follow, both being parts of the
same twitter social graph, have the largest percentage of
``superstar'' nodes compared to the other networks.

\section{Graph Partitioning}

GraphX is a graph analytics framework built on Apache Spark.  Spark
uses the abstraction of Resilient Distributed Datasets (RDDs) to
structure, sequence, and schedule MapReduce computations over
partitioned data.  GraphX uses RDD partitions to store graph
partitions in its own representation, and maps computations expressed
in the Bulk-Synchronous Parallel (BSP) programming model to
lower-level RDD MapReduce operations.


GraphX uses vertex cut partitioning; it first distributes graph edges
into RDD partitions, and then builds a graph partition representation,
local to each RDD partition, containing local and replicated vertices
as well as metadata describing all necessary communication to
implement a BSP computation.  GraphX includes four ways to initially
partition the edge list that represents each GraphX graph, which
result into four different vertex cut strategies.  In addition to
these, we develop and use two additional partitioning strategies
aiming to explore the design space and optimize for different metrics.
Specifically, we use the following GraphX partitioners:
\\
\textbf{Random Vertex Cut (RVC)} assigns edges to partitions by
hashing together the source and destination vertex IDs, resulting in a
random vertex cut that collocates all same-direction edges between two
vertices.
\\
\textbf{Edge Partition 1D (1D)} assigns edges to partitions by hashing
the source vertex ID.This causes all edges with the same source vertex
to be collocated in the same partition.
\\
\textbf{Edge Partition 2D (2D)} arranges all partitions into a square
matrix and picks the column on the basis of the source vertex's hash
and the row on the basis of the destination vertex's hash.  This
strategy guarantees a $2 ∗ \sqrt{N}$ upper bound on vertex replication
where $N$ is the number of partitions.  Moreover, this strategy works
best if the number of partitions is a perfect square; if not, the
algorithm uses the next largest number and potentially creates
imbalanced partitioning.
\\
\textbf{Canonical Random Vertex Cut (CRVC)} assigns edges to
partitions by hashing the source and destination vertex IDs in a
canonical direction, resulting in a random vertex cut that collocates
all edges between two vertices, regardless of direction. For example
$(u, v)$ and $(v, u)$ hash to the same partition in Canonical Random
Vertex Cut but not necessarily under RVC.

To explore the design space further, we designed and implemented two
additional partitioning algorithms by changing some of the assumptions
in GraphX partitioners:
\\
\textbf{Source Cut (SC):} assign edges to partition by simple modulo
of the the source vertex IDs.  This is almost equivalent to the 1D
partitioner, but also assuming that vertex IDs may capture a metric of
locality.  We expected this partitioner to result into less balanced
partitions, as the hashing function in 1D achieves a more uniform
distribution, but in cases where vertex ID similarity captures
locality, we expected this partitioner to take advantage of it.
\\
\textbf{Destination Cut (DC):} assigns edges to partitions by simple
modulo on only the destination vertex IDs.  This is similar to SC
except it uses the vertex ID of the edge destination to assign edges
to partitions. As in SC, we expect any correlation between vertex IDs
and locality to be captured, at the expense of load-balancing.

\subsection{Characterization Metrics}

For each of these partitioners we measure a set of metrics that
capture the properties of the partitioning, and help understand how a
partitioner works on each different dataset.  We use a set of standard
metrics that have been used to predict performance~\cite{metrics},
augmenting the set with the standard deviation of the edge-partition
sizes, the number of vertices that are replicated in other partitions,
and the number of vertices that are not replicated and only reside in
a single partition.  Note that even if GraphX uses vertex cut
partitioning, it does not store solely edges on each partition, but
instead reconstructs the vertices per edge partition, and finally
creates a data structure that includes the partition's vertex list.
\\
\textbf{Balance} aims to capture how balanced partition sizes are, and
is equal to the ratio of the number of edges in the biggest partition,
over the average number of edges per partition.
\\
\textbf{Non-Cut} describes the number of vertices that are not
replicated among partitions and reside in a single partition.
\\
\textbf{Cut} is the number of vertices that exist in more than one
partition, irrespective of how many copies of each cut vertex there
are.  In essence, these are the vertices who are copied across
partitions into the number of copies that forms the Communication Cost
metric.
\\
\textbf{Communication Cost} (CommCost) aims to approximate the
communication cost incurred by the partitioning for a hypothetical
Bulk-Synchronous Parallel computation that stores a fixed-sized state
on all vertices.  It is equal to the total number of copies of
replicated vertices that exist in more than one partition, as this is
the number of messages that need to be exchanged on every superstep to
agree on the state stored in these vertices.
\\
\textbf{Edge Partition Standard Deviation} (PartStDev) is the standard
deviation of the number of edges per partition.  Similarly to Balance,
it constitutes a measure of imbalance in the edge partitions.  Note,
however, that imbalance in edge partitions does not necessarily mean
imbalanced usage of memory, as the final partitions also hold the
vertices of all included edges.

Note that some of the metrics are related, since the sum of
Communication Cost plus Non-Cut Vertices is always equal to the sum of
Vertices to Same plus Vertices to Other.  This is because they
correspond to different breakdowns of the total number of vertex
replicas that exist over all partitions, one is based on the existence
of referring edges in the same partition, and the other is based on
the non existence of referring edges in other partitions.

\subsection{Social Network Analytics Algorithms}

To measure the effect of the differences in partitioning presented
above, we ran the following graph algorithms in GraphX.
\\
\textbf{PageRank (PR)} computes the importance of websites within the
web graph, based on the shape of the graph around it.
\\
\textbf{Connected Components (CC)} computes the number of connected
components of the graph.  The connected components algorithm labels
each connected component of the graph with ID of it's lowest-numbered
vertex.
\\
\textbf{Triangles (TR)} computes the number of triangles passing
through each vertex and sums to find the number for the whole graph.
\\
\textbf{Shortest Path (SSSP)} computes shortest paths to the given set
of landmark vertices, returns a graph where each vertex attribute is a
map containing the shortest-path distance to each reachable landmark.

\section{Evaluation}

We ran all experiments on a cluster of 5 Intel(R) Xeon(R) E5-2630 CPUs
with 256GB of main memory, configured as 1 Spark Driver and 4 Spark
Executors. Each Executor used 220GB of memory and 32 cores, resulting
into 128 total cores.  For algorithms that iterate either a number of
times or to fixpoint, namely PageRank and Connected Components, we run
each experiment for 10 iterations.  All times reported are the average
of 5 runs for each of two granularity configurations: Configuration
(i) uses 128 partitions and (ii) uses 256 partitions.
We measured all metrics for all configurations, datasets and
partitioners, and computed their correlation with running
time\footnote{Due to space constraints, we refer the reader to the
accompanying Technical Report for a presentation of the results on all
metrics, and focus only on the most important findings in this
paper.}.

\begin{figure}
    \begin{minipage}{0.48\textwidth}
        \centering
        \includegraphics[width=\textwidth]{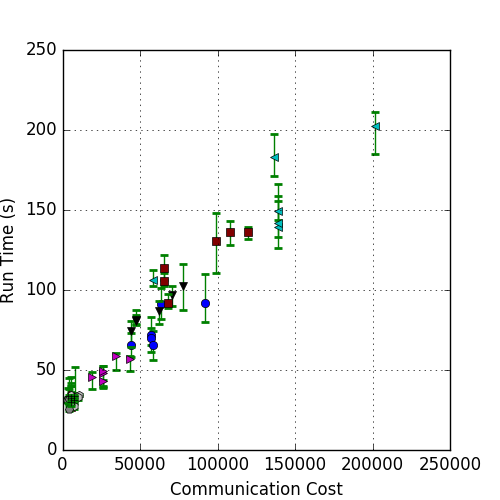}
        Configuration (i)
    \end{minipage}
    \hfill
    \begin{minipage}{0.48\textwidth}
        \centering
        \includegraphics[width=\textwidth]{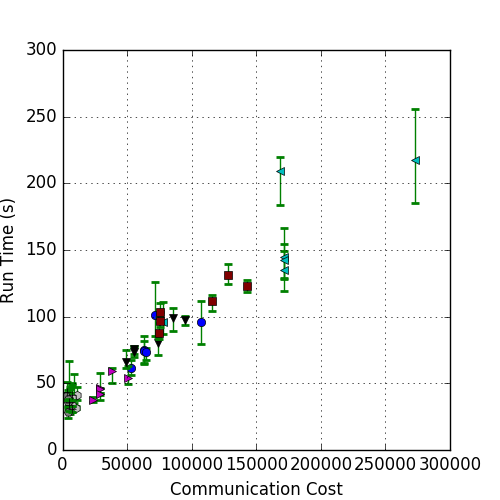}
        Configuration (ii)
    \end{minipage}
    \hfill
    \begin{minipage}{\textwidth}
        \centering
        \includegraphics[width=\textwidth]{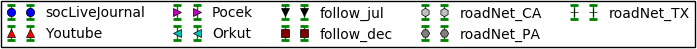}
        Legend (omitted from subsequent Figures)
    \end{minipage}
    \caption{Correlation between execution time and communication cost for PageRank}
    \label{fig:pagerank}
\end{figure}

Figure~\ref{fig:pagerank} shows the correlation between execution time
and the communication cost for PageRank. As expected, we found the
Communication Cost metric to be consistently the most important
predictor of execution time for algorithms similar to PageRank, where
computation per node is small compared to the exchanged messages, with
correlation coefficients of 95\% and 96\% for partitionings (i) and
(ii) respectively.  For PageRank, finer grain partitioning increases
the execution time even for the largest dataset, as the algorithm is
communication bound.
We found that the number of partitions affects not only the execution
time but also the optimal partition strategy. For the coarse-grain
configuration (i), the best partition strategy for follow-jul,
follow-dec and YouTube is 2D, while for all others it is DC.  For the
fine-grain configuration (ii), the best partition strategy for YouTube
,RoadNet-PA, RoadNet-TX is DC and for all others it is 2D.  In
general, we found best to opt for DC for smaller datasets and 2D for
large datasets, as even in the one exception to this rule, for YouTube
the two partitioners differed only very slightly.  Both of these
partitioners aim to optimize for communication cost, with 2D achieving
better locality on large datasets.

\begin{figure}[t]
\begin{minipage}{0.48\textwidth}
  \centering
  \includegraphics[width=\textwidth]{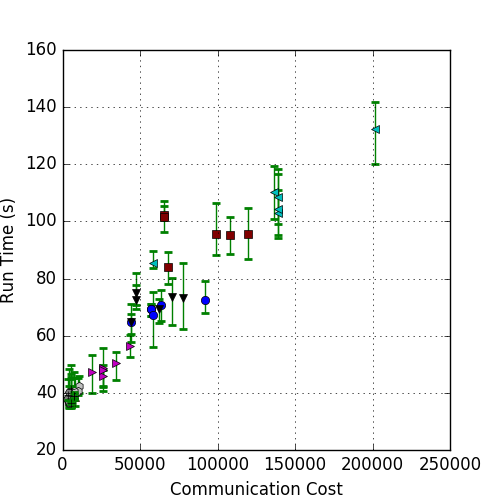}
  Configuration (i)
\end{minipage}
\hfill
\begin{minipage}{0.48\textwidth}
  \centering
  \includegraphics[width=\textwidth]{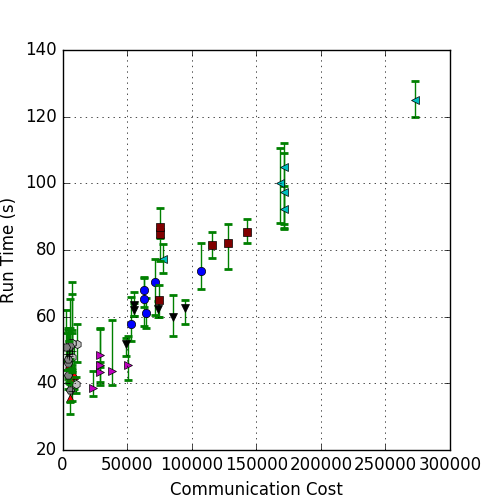}
  Configuration (ii)
\end{minipage}
\caption{Correlation between execution time and communication cost for Connected Components}
\label{fig:cc}
\end{figure}

Figure~\ref{fig:cc} shows the correlation between execution time and
communication cost for Connected Components.  CC is a
label-propagation algorithm where the values of most vertices converge
very fast and will not be updated in subsequent iterations.  As in
PageRank, Communication Cost is the best performance predictor for
both configurations, correlated at 92\% and 94\% respectively.  In
comparison to PageRank, the CC algorithm is less consistently
dominated by Communication Cost, however; after a few iterations the
algorithm converges for most vertices, allowing the fine-grain
configuration (ii) to perform better for all but the smallest
datasets, compared to the coarse-grained configuration (i).
For configuration (i), the best partition strategy for follow-jul,
follow-dec, Orkut and socLiveJournal is 2D. For the five smaller
datasets, Pocek, YouTube, RoadNet-CA, RoadNet-PA and RoadNet-TX the
best partition strategy is 1D, although the difference is in the
noise.  On configuration (ii), the best partition strategy for all
datasets is 2D.  Execution time from configuration (i) to (ii)
decreases up to 22\% on the bigger datasets, because after the first
few iterations not all vertices need to be revisited. This results in
partitions of similar size being load-imbalanced with respect to
running time, causing the fine-grain configuration (ii) to perform
better.

\begin{figure}[t]
\begin{minipage}{0.48\textwidth}
    \centering
    \includegraphics[width=\textwidth]{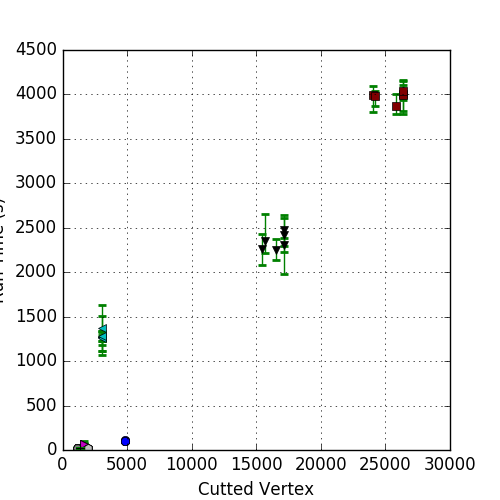}
    Configuration (i)
\end{minipage}
\hfill
\begin{minipage}{0.48\textwidth}
    \centering
    \includegraphics[width=\textwidth]{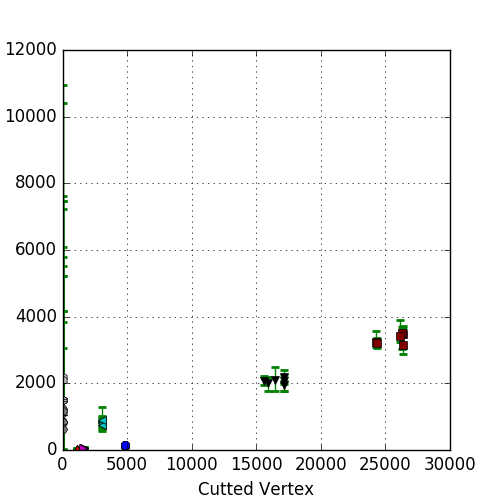}
    Configuration (ii)
\end{minipage}
\caption{Correlation between Execution Time and Cut-Vertices for Triangle Count}
    \label{fig:triangles}
\end{figure}

Figure~\ref{fig:triangles} shows the correlation between execution
time and the number of Cut vertices for Triangle Count (TR).  TR
performs much more computation per node compared to PageRank and CC,
and much less communication.  We found the most correlated
partitioning metric to execution time in this case to be the number of
vertices replicated across more than one partitions, as these incur
additional reductions to communicate per-vertex state in GraphX (and
all Pregel-like systems).  Correlation to execution time was 95\% and
97\% for configurations (i) and (ii), respectively.  Interestingly,
the Communication Cost metric correlation coefficient is only 43\% and
34\%, respectively.
For configuration (i), the best partitioning strategy for follow-jul,
follow-dec, RoadNet-PA and RoadNet-TX is DC, for Orkut and YouTube it
is SC, for Pocek and Roadnet-CA it is 2D and for socLiveJournal it is
CRVC.  As seen from Figure~\ref{fig:triangles}, however, none of the
partitioners manages to optimize for this metric much better than the
rest, resulting in performance differences of 5-10\% between the best
and worst partitioners in all datasets except the smallest.  For
configuration (ii), the best partitioning strategy for follow-jul and
YouTube is 2D, for follow-dec, Orkut, RoadNet-CA, RoadNet-PA,
RoadNet-TX, and socLiveJournal it is CRVC, and for Pocek it is DC,
with most differences being in the noise.
With respect to granularity, fine-grain configuration (ii) outperforms
configuration (i) consistently, by up to 40\% for Orkut and 20\% for
follow-dec.


\begin{figure}
\begin{minipage}{0.48\textwidth}
  \centering
  \includegraphics[width=\textwidth]{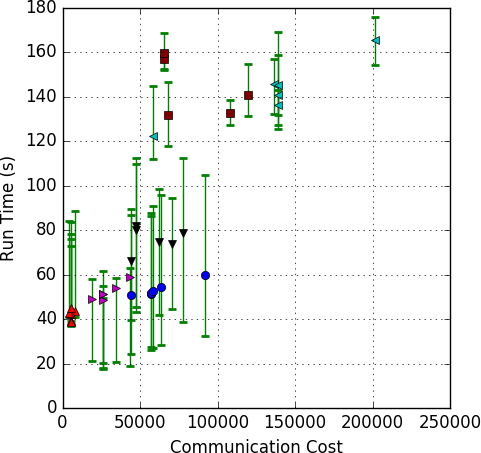}
  Configuration (i)
\end{minipage}
\hfill
\begin{minipage}{0.48\textwidth}
  \centering
  \includegraphics[width=\textwidth]{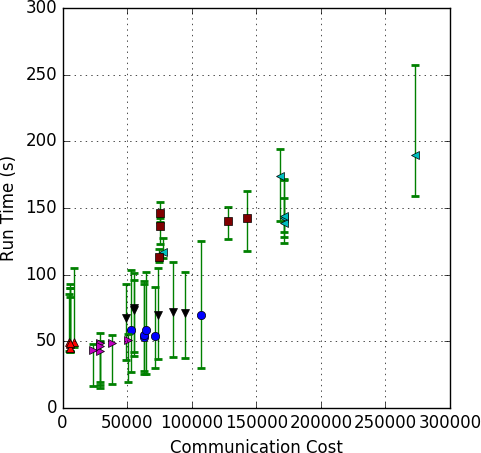}
  Configuration (ii)
\end{minipage}
\caption{Correlation between Execution Time and Communication Cost for SSSP}
\label{fig:spath}
\end{figure}

To evaluate the SSSP algorithm, as it is highly sensitive to the
single vertex selected as the shortest path source, we randomly
selected 5 source vertices in each dataset and measured all
partitioners for each of the source vertices.  This way we aim to
average over both a number of runs and a number of possible source
vertices, as the algorithm may be highly sensitive to the shape and
density of each dataset as well as the source vertex.

Figure~\ref{fig:spath} shows the correlation between execution time
and Communication Cost for SSSP.  As with PageRank, SSSP is highly
communication bound, and may require a number of iterations related to
the graph diameter to finish.  The correlation to Communication Cost
for configurations (i) and (ii) is 80\% and 86\% respectively.  On
configuration (i), the best partition strategy is 2D for Follow,
Follow\_dec, Orkut and socLiveJournal, for Pocek it is 1D, and for
YouTube it is SC. On configuration (ii), the best partition strategy
is 2D for Follow, Follow\_dec, Orkut and YouTube, and for Pocek and
socLiveJournal it is 1D.  Granularity seems to not affect the
execution time consistently in the case of SSSP.
The grid datasets (Roadnet-CA, RoadNet-PA and RoadNet-TX) are not
shown in the plot, as Spark did not complete SSSP, due to out of
memory errors.

Due to the setup of this experiment, the SSSP measurements exhibit a
much greater variance compared to the rest of the algorithms.  This is
because the average-of-5 results depicted correspond to 5 different
source vertices, not to 5 exactly identical executions as in the other
algorithms.

Overall, we found that for algorithms with complexity mostly related
to the number of edges should prefer partitioners that optimize the
Communication Cost metric; for algorithms whose complexity mostly
depends on the number of vertices, partitioners should be compared
using the Cut Vertices metric, as a better approximation of the
communication overhead of each Pregel/BSP superstep.

We performed two additional experiments to better understand the
impact of actual communication costs on GraphX, by changing the
physical network and storage resources available in our Spark cluster.
We ran the two additional experiments based on the configuration (ii).
Specifically, we ran the PageRank algorithm on the big dataset,
follow\_dec after updating the network Infrastructure to 40Gbps,
compared to 1Gbps above.  We set up two new configurations:
configuration (iii) uses HDFS on Hard Disk Drives for storage, as in
configuration (ii); and configuration (iv) uses local Solid State
Drives on every executor machine.  On each configuration we take the
average of five different runs and compare with the configuration (ii)
average. We found that execution time is on average 15\% less for
configuration (iii) and 20\% less for configuration (iv). This shows
that selecting a good partitioner has a bigger impact on performance
for better infrastructure.

\section{Related Work}

Apache GraphX implements the Bulk Synchronous Processing (BSP)
model~\cite{BSPmodel} for graph processing.  BSP was first used for
large-scale graph analytics in Pregel~\cite{pregelSystem} introduced
by Malecwiz et al. Except for GraphX, there are several other
Pregel-like systems for graph analytics, such as Giraph,
GPS~\cite{GPS}, Mizan~\cite{mizan} and GraphLab.  As these graph
analytics frameworks aim to be generic and support any algorithm, they
are often forced to generalize their design over all graph
computations, resulting in sub-optimal performance compared to a
hand-crafted implementation of each algorithm.  Satish et
al.~\cite{satish} demonstrate a huge performance gap between all the
state-of-the-art graph processing frameworks and the best hand-crafted
implementation, which they call the ``ninja gap.''

A lot of related work focuses on the performance comparison of these
systems, without however producing a common consensus. Han et al.,
provide an experimental comparison of Pregel-like
systems~\cite{comparisonOfPregelLikeSystems} of Giraph, GPS, Mizan and
GraphLab.  They conclude that Giraph and GraphLab have better
performance than the other two frameworks.  Ching et
al.~\cite{oneTrillionEdges} provide a comparison between GraphX and a
custom version of Giraph, and found that Apache Giraph does not
outperform GraphX.  Verma et al.~\cite{comparisonPartStrat} present an
experimental comparison between partition strategies in Distributed
Graph Processing on different platforms on PowerGraph, PowerLyra and
GraphX.  Through their experiments they prove that no single
partitioning is the best to fit overall situations and give a
heuristic guide for selecting a partitioner algorithm for PowerGraph
and PowerLyra. We have improved on these findings by studying the
effect on much larger graphs, the effect of partitioning granularity,
and explained the performance difference observed in terms of
edge-based or vertex-based communication metrics.

There are several approaches to graph partitioning in the literature,
aiming to optimize the performance of graph processing frameworks.
Fennel~\cite{fennel} is a one-pass streaming graph partitioning
algorithm achieving significant performance improvement than previous
implementations.  Stanton et al.~\cite{streamingGraphPartitioning}
present a streaming graph partitioner that eliminates communication
cost of partitioning by partitioning as the graph is loaded.  Karypis
and Kumar~\cite{karypis1998fast} have proposed hierarchical
partitioning similar to clustering and community detection
computations, to optimize communication costs.

\section{Conclusions}
Graph Analytics computations are complex and highly dependent on the
properties of each specific dataset.  Many such applications use
standard analytics runtimes, optimized for the general case, even
though social network datasets and computations have very particular
characteristics.  In this work we investigate how a computation can be
better optimized for social network datasets by tailoring the
partitioning strategy to the dataset and to the computation.

We measure the effect of partitioning on performance for many
datasets, partitioning strategies, and analytics algorithms. Over all
partitioners, we found that communication cost not affect the
performance in all cases except in case of algorithms keeping a lot of
per-vertex state and per-vertex computation, such as Triangle Count.
We show that granularity plays a significant role in performance, and
provide heuristics as to selecting the partitioning granularity based
on the dataset size and algorithm characteristics.

\bibliographystyle{abbrv}
\bibliography{techReport} 

\begin{thebibliography}{10}

\bibitem{karypis}
A.~Abou-Rjeili and G.~Karypis.
\newblock Multilevel algorithms for partitioning power-law graphs.
\newblock In {\em International Conference on Parallel and Distributed
  Processing (IPDPS)}, pages 124--124, 2006.

\bibitem{giraph}
C.~Avery.
\newblock Giraph: Large-scale graph processing infrastructure on hadoop.
\newblock {\em Hadoop Summit. Santa Clara}, 11(3):5--9, 2011.

\bibitem{backstrom2006group}
L.~Backstrom, D.~Huttenlocher, J.~Kleinberg, and X.~Lan.
\newblock Group formation in large social networks: membership, growth, and
  evolution.
\newblock In {\em ACM SIGKDD international conference on Knowledge discovery
  and data mining}, pages 44--54, 2006.

\bibitem{oneTrillionEdges}
A.~Ching, S.~Edunov, M.~Kabiljo, D.~Logothetis, and S.~Muthukrishnan.
\newblock One trillion edges: Graph processing at facebook-scale.
\newblock {\em Proc. VLDB Endow.}, 8(12):1804--1815, Aug. 2015.

\bibitem{mapReduce}
J.~Dean and S.~Ghemawat.
\newblock Mapreduce: Simplified data processing on large clusters.
\newblock {\em Commun. ACM}, 51(1):107--113, Jan. 2008.

\bibitem{powergraph}
J.~E. Gonzalez, Y.~Low, H.~Gu, D.~Bickson, and C.~Guestrin.
\newblock Powergraph: Distributed graph-parallel computation on natural graphs.
\newblock In {\em OSDI}, page~2, 2012.

\bibitem{comparisonOfPregelLikeSystems}
M.~Han, K.~Daudjee, K.~Ammar, M.~T. \"{O}zsu, X.~Wang, and T.~Jin.
\newblock An experimental comparison of pregel-like graph processing systems.
\newblock {\em Proc. VLDB Endow.}, 7(12):1047--1058, Aug. 2014.

\bibitem{karypis1998fast}
G.~Karypis and V.~Kumar.
\newblock A fast and high quality multilevel scheme for partitioning irregular
  graphs.
\newblock {\em SIAM Journal on scientific Computing}, 20(1):359--392, 1998.

\bibitem{mizan}
Z.~Khayyat, K.~Awara, A.~Alonazi, H.~Jamjoom, D.~Williams, and P.~Kalnis.
\newblock Mizan: A system for dynamic load balancing in large-scale graph
  processing.
\newblock In {\em ACM European Conference on Computer Systems (EuroSys '13)},
  pages 169--182, 2013.

\bibitem{graphChi}
A.~Kyrola, G.~Blelloch, and C.~Guestrin.
\newblock Graphchi: Large-scale graph computation on just a pc.
\newblock In {\em USENIX Conference on Operating Systems Design and
  Implementation (OSDI)}, pages 31--46, 2012.

\bibitem{snap}
J.~Leskovec and A.~Krevl.
\newblock {SNAP Datasets}: {Stanford} large network dataset collection.
\newblock \url{http://snap.stanford.edu/data}, June 2014.

\bibitem{road}
J.~Leskovec, K.~J. Lang, A.~Dasgupta, and M.~W. Mahoney.
\newblock Community structure in large networks: Natural cluster sizes and the
  absence of large well-defined clusters.
\newblock {\em CoRR}, abs/0810.1355, 2008.

\bibitem{pregelSystem}
G.~Malewicz, M.~H. Austern, A.~J. Bik, J.~C. Dehnert, I.~Horn, N.~Leiser, and
  G.~Czajkowski.
\newblock Pregel: A system for large-scale graph processing.
\newblock In {\em ACM SIGMOD International Conference on Management of Data},
  pages 135--146, 2010.

\bibitem{metrics}
H.~Mykhailenko, G.~Neglia, and F.~Huet.
\newblock Which metrics for vertex-cut partitioning?
\newblock In {\em Internet Technology and Secured Transactions (ICITST)}, pages
  74--79, 2016.

\bibitem{galois}
K.~Pingali, D.~Nguyen, M.~Kulkarni, M.~Burtscher, M.~A. Hassaan, R.~Kaleem,
  T.-H. Lee, A.~Lenharth, R.~Manevich, M.~M{\'e}ndez-Lojo, et~al.
\newblock The tao of parallelism in algorithms.
\newblock In {\em ACM Sigplan Notices}, pages 12--25, 2011.

\bibitem{rahimian}
F.~Rahimian, A.~H. Payberah, S.~Girdzijauskas, and S.~Haridi.
\newblock Distributed vertex-cut partitioning.
\newblock In {\em IFIP International Conference on Distributed Applications and
  Interoperable Systems}, pages 186--200, 2014.

\bibitem{GPS}
S.~Salihoglu and J.~Widom.
\newblock Gps: A graph processing system.
\newblock In {\em International Conference on Scientific and Statistical
  Database Management (SSDBM)}, pages 22:1--22:12, 2013.

\bibitem{satish}
N.~Satish, N.~Sundaram, M.~M.~A. Patwary, J.~Seo, J.~Park, M.~A. Hassaan,
  S.~Sengupta, Z.~Yin, and P.~Dubey.
\newblock Navigating the maze of graph analytics frameworks using massive graph
  datasets.
\newblock In {\em ACM SIGMOD international conference on Management of data},
  pages 979--990, 2014.

\bibitem{streamingGraphPartitioning}
I.~Stanton and G.~Kliot.
\newblock Streaming graph partitioning for large distributed graphs.
\newblock In {\em Proceedings of the 18th ACM SIGKDD International Conference
  on Knowledge Discovery and Data Mining}, KDD '12, pages 1222--1230, New York,
  NY, USA, 2012. ACM.

\bibitem{pocekdata}
L.~Takac and M.~Zabovsky.
\newblock Data analysis in public social networks.
\newblock {\em International Scientific Conference \& International Workshop
  Present Day Trends of Innovations 2012}, 2012.

\bibitem{fennel}
C.~Tsourakakis, C.~Gkantsidis, B.~Radunovic, and M.~Vojnovic.
\newblock Fennel: Streaming graph partitioning for massive scale graphs.
\newblock In {\em Proceedings of the 7th ACM International Conference on Web
  Search and Data Mining}, WSDM '14, pages 333--342, New York, NY, USA, 2014.
  ACM.

\bibitem{BSPmodel}
L.~G. Valiant.
\newblock A bridging model for parallel computation.
\newblock {\em Commun. ACM}, 33(8):103--111, Aug. 1990.

\bibitem{comparisonPartStrat}
S.~Verma, L.~M. Leslie, Y.~Shin, and I.~Gupta.
\newblock An experimental comparison of partitioning strategies in distributed
  graph processing.
\newblock {\em Proc. VLDB Endow.}, 10(5):493--504, Jan. 2017.

\bibitem{graphx}
R.~S. Xin, J.~E. Gonzalez, M.~J. Franklin, and I.~Stoica.
\newblock Graphx: A resilient distributed graph system on spark.
\newblock In {\em International Workshop on Graph Data Management Experiences
  and Systems}, page~2, 2013.

\bibitem{youtubedata}
J.~Yang and J.~Leskovec.
\newblock Defining and evaluating network communities based on ground-truth.
\newblock {\em Knowledge and Information Systems}, 42(1):181--213, 2015.

\bibitem{graphlab}
L.~Yucheng, G.~Joseph, K.~Aapo, B.~Danny, G.~Carlos, and M.~Joseph.
\newblock Graphlab: A new framework for parallel machine learning.
\newblock In {\em Proc. Int’l Conf. Uncertainty in Artificial Intelligence
  (UAI’10)}, 2010.

\bibitem{spark}
M.~Zaharia, M.~Chowdhury, M.~J. Franklin, S.~Shenker, and I.~Stoica.
\newblock Spark: Cluster computing with working sets.
\newblock In {\em HotCloud 2010}, 2010.

\end{thebibliography}

\begin{appendices}
\section{Partitioning Metrics}

Table~\ref{tab:128metrics} presents a characterization of the
different partitioning algorithms using all metrics over all datasets,
for 128 partitions.  Note that different partitioners can produce very
different metrics.  For instance, note that the Random Vertex Cut
partitioner consistently produces partitionings where a very small
number of vertices are not Cut, i.e., replicated in more than one
partition.  On the other hand, in one case the 2D partitioner has 0
vertices that are not replicated.  Also, the CRVC and RVC partitioners
achieve more balanced partitions, possibly at the cost of the expected
communication between partitions.

\begin{table*}
\begin{center}
\resizebox{0.76\textwidth}{!}{
\begin{tabular}{|c|c|r|r|r|r|r|}
    \hline
    Dataset & Partitioner & Balance & NonCut & Cut & CommCost & PartStDev \\ \hline
    \hline
    \multirow{6}{*}{\parbox[t]{2mm}{\rotatebox[origin=c]{90}{RoadNet-PA}}} 
    &RVC    &1.01   & 26        &1,088,066      &6,039,312      &146.05   \\ \cline{2-7}
    &1D     &1.00   & 10        &1,088,082      &4,151,717      & 84.20   \\ \cline{2-7}
    &2D     &1.01   & 21        &1,088,071      &5,696,695      &139.85   \\ \cline{2-7}
    &CRVC   &1.01   & 19        &1,088,073      &3,057,457      &195.00   \\ \cline{2-7}
    &SC     &1.00   & 20        &1,088,072      &4,151,995      & 78.99   \\ \cline{2-7}
    &DC     &1.00   & 20        &1,088,072      &4,151,995      & 78.99   \\
    \hline

    \multirow{6}{*}{\parbox[t]{2mm}{\rotatebox[origin=c]{90}{YouTube}}} 
    &RVC    &1.01   &    64     &1,134,826      &8,104,655      &  196.38   \\ \cline{2-7}
    &1D     &1.68   &14,993     &1,119,897      &5,611,447      &4,652.58   \\ \cline{2-7}
    &2D     &1.13   &   425     &1,134,465      &5,522,410      &2,176.41   \\ \cline{2-7}
    &CRVC   &1.01   &    51     &1,134,839      &4,504,403      &  308.78   \\ \cline{2-7}
    &SC     &1.59   &10,662     &1,124,228      &5,616,752      &4,436.65   \\ \cline{2-7}
    &DC     &1.59   &10,662     &1,124,228      &5,616,752      &4,436.65   \\
    \hline

    \multirow{6}{*}{\parbox[t]{2mm}{\rotatebox[origin=c]{90}{RoadNet-TX}}} 
    &RVC    &1.01   & 91        &1,379,826      &7,529,594      &180.93     \\ \cline{2-7}
    &1D     &1.00   & 55        &1,379,862      &5,199,911      & 95.45     \\ \cline{2-7}
    &2D     &1.01   & 82        &1,379,835      &7,106,117      &153.36     \\ \cline{2-7}
    &CRVC   &1.01   &  6        &1,379,911      &3,811,064      &254.95     \\ \cline{2-7}
    &SC     &1.00   & 61        &1,379,856      &5,199,970      & 92.88     \\ \cline{2-7}
    &DC     &1.00   & 61        &1,379,856      &5,199,970      & 92.88     \\
    \hline

    \multirow{6}{*}{\parbox[t]{2mm}{\rotatebox[origin=c]{90}{Pokec}}} 
    &RVC    &1.00   &  625      &1,632,178      &43,452,217     &  491.73   \\ \cline{2-7}
    &1D     &1.03   &1,338      &1,631,465      &26,225,882     &2,873.98   \\ \cline{2-7}
    &2D     &1.01   &  486      &1,632,317      &18,820,634     &1,137.80   \\ \cline{2-7}
    &CRVC   &1.00   &  262      &1,632,541      &34,418,293     &  612.12   \\ \cline{2-7}
    &SC     &1.05   &2,471      &1,630,332      &26,222,846     &2,978.84   \\ \cline{2-7}
    &DC     &1.07   &5,202      &1,627,601      &26,294,017     &3,457.81   \\
    \hline

    \multirow{6}{*}{\parbox[t]{2mm}{\rotatebox[origin=c]{90}{Roadnet-CA}}} 
    &RVC    &1.01   &118        &1,965,088      &10,838,867     &196.95   \\ \cline{2-7}
    &1D     &1.00   & 40        &1,965,166      & 7,465,034     &109.38   \\ \cline{2-7}
    &2D     &1.00   & 13        &1,965,193      &10,252,103     &173.70   \\ \cline{2-7}
    &CRVC   &1.01   & 55        &1,965,151      & 5,486,526     &292.22   \\ \cline{2-7}
    &SC     &1.00   & 29        &1,965,177      & 7,465,363     &120.08   \\ \cline{2-7}
    &DC     &1.00   & 29        &1,965,177      & 7,465,363     &120.08   \\
    \hline

    \multirow{6}{*}{\parbox[t]{2mm}{\rotatebox[origin=c]{90}{Orkut}}} 
    &RVC    &1.00   &   48      &3,072,393      &201,579,891    & 1,388.70  \\ \cline{2-7}
    &1D     &1.03   &5,134      &3,067,307      &139,337,672    &22,320.95  \\ \cline{2-7}
    &2D     &1.01   &   13      &3,072,428      & 58,110,415    & 8,115.76  \\ \cline{2-7}
    &CRVC   &1.00   &  206      &3,072,235      &136,670,436    & 1,944.62  \\ \cline{2-7}
    &SC     &1.03   &1,384      &3,071,057      &139,339,199    &22,968.71  \\ \cline{2-7}
    &DC     &1.03   &1,384      &3,071,057      &139,339,199    &22,968.71  \\
    \hline

    \multirow{6}{*}{\parbox[t]{2mm}{\rotatebox[origin=c]{90}{socLiveJournal}}} 
    &RVC    &1.00   &   16      &4,847,555      &91,727,814     & 732.58    \\ \cline{2-7}
    &1D     &1.04   &6,967      &4,840,604      &57,194,270     &6375.69    \\ \cline{2-7}
    &2D     &1.01   &  180      &4,847,391      &44,408,628     &2891.27    \\ \cline{2-7}
    &CRVC   &1.00   &  374      &4,847,197      &63,830,074     &1021.94    \\ \cline{2-7}
    &SC     &1.04   &9,866      &4,837,705      &57,208,429     &5973.63    \\ \cline{2-7} 
    &DC     &1.04   &2,896      &4,844,675      &58,398,861     &7245.30    \\
    \hline

    \multirow{6}{*}{\parbox[t]{2mm}{\rotatebox[origin=c]{90}{Follow\_jul}}} 
    &RVC    &1.00   &       12  &17,172,130     &77,909,024     &    943.02 \\ \cline{2-7}
    &1D     &8.59   &1,743,865  &15,428,277     &47,218,160     &743,525.59 \\ \cline{2-7}
    &2D     &1.72   &   13,166  &17,158,976     &44,263,566     &201,600.81 \\ \cline{2-7}
    &CRVC   &1.00   &      206  &17,171,936     &70,607,064     &   1171.01 \\ \cline{2-7}
    &SC     &8.59   &1,471,524  &15,700,618     &47,441,754     &756,719.19 \\ \cline{2-7}
    &DC     &4.26   &  623,712  &16,548,430     &61,962,146     &344,698.67 \\
    \hline

    \multirow{6}{*}{\parbox[t]{2mm}{\rotatebox[origin=c]{90}{Follow\_dec}}} 
    &RVC    &1.00   &       66  &26,339,905     &119,503,790    &    1,134.83 \\ \cline{2-7}
    &1D     &10.04  &2,282,728  &24,057,243     & 65,446,694    &1,333,180.21 \\ \cline{2-7}
    &2D     &2.04   &    9,140  &26,330,831     & 67,832,243    &  377,868.97 \\ \cline{2-7}
    &CRVC   &1.00   &      717  &26,339,254     &107,795,617    &    1,403.23 \\ \cline{2-7}
    &SC     &10.04  &2,161,293  &24,178,678     & 65,517,506    &1,364,952.64 \\ \cline{2-7}
    &DC     &4.86   &  525,799  &25,814,172     & 98,839,823    &  749,972.78 \\
    \hline

\end{tabular}
}
\end{center}
\caption{Partitioning metrics for 128 partitions}
\label{tab:128metrics}
\end{table*}

Table \ref{tab:256metrics} presents the same characterization of all
partitioning algorithms for 256 partitions.  Note that, in comparison
to the corresponding metrics for 128 partitions, communication costs
increase as expected, but are significantly lower than double.
Moreover, splitting the networks into 256 partitions achieves much
higher balance factor, as the biggest partition produced is a lot
bigger than the smaller, for more partitions.

\begin{table*}
\begin{center}
\resizebox{0.76\textwidth}{!}{
\begin{tabular}{|c|c|r|r|r|r|r|}
    \hline
    Dataset & Partitioner & Balance & NonCut & Cut & CommCost & PartStDev \\ \hline
    \hline
    \multirow{6}{*}{\parbox[t]{2mm}{\rotatebox[origin=c]{90}{RoadNet-PA}}} 
    &RVC    &1.02   &17          &1,088,075    & 6,102,980   &   108.69     \\ \cline{2-7}
    &1D     &1.01   &3           &1,088,089    & 4,162,945   &    59.24     \\ \cline{2-7}
    &2D     &2.61   &0           &1,088,092    & 5,789,569   & 7,663.64     \\ \cline{2-7}
    &CRVC   &1.02   &5           &1,088,087    & 3,070,630   &   149.07     \\ \cline{2-7}
    &SC     &1.01   &16          &1,088,076    & 4,163,198   &    58.84     \\ \cline{2-7}
    &DC     &1.01   &16          &1,088,076    & 4,163,198   &    58.84     \\
    \hline

    \multirow{6}{*}{\parbox[t]{2mm}{\rotatebox[origin=c]{90}{YouTube}}} 
    &RVC    &1.01   &    64     &1,134,826    &8,104,655     &  196.38     \\ \cline{2-7}
    &1D     &1.68   &14,993     &1,119,897    &5,611,447     &4,652.58     \\ \cline{2-7}
    &2D     &1.13   &   425     &1,134,465    &5,522,410     &2,176.41     \\ \cline{2-7}
    &CRVC   &1.01   &    51     &1,134,839    &4,504,403     &  308.78     \\ \cline{2-7}
    &SC     &1.59   &10,662     &1,124,228    &5,616,752     &4,436.65     \\ \cline{2-7}
    &DC     &1.59   &10,662     &1,124,228    &5,616,752     &4,436.65     \\
    \hline

    \multirow{6}{*}{\parbox[t]{2mm}{\rotatebox[origin=c]{90}{RoadNet-TX}}} 
    &RVC    &1.02   &7          &1,379,910    & 7,608,152    &   124.73     \\ \cline{2-7}
    &1D     &1.01   &12         &1,379,905    & 5,213,392    &    68.96     \\ \cline{2-7}
    &2D     &2.53   &0          &1,379,917    & 7,217,760    & 9,178.92     \\ \cline{2-7}
    &CRVC   &1.03   &12         &1,379,905    & 3,827,233    &   171.65     \\ \cline{2-7}
    &SC     &1.01   &33         &1,379,884    & 5,213,644    &    68.46     \\ \cline{2-7}
    &DC     &1.01   &33         &1,379,884    & 5,213,644    &    68.46     \\
    \hline

    \multirow{6}{*}{\parbox[t]{2mm}{\rotatebox[origin=c]{90}{Pocek}}} 
    &RVC    &1.00   &  625      &1,632,178    &43,452,217    &  491.73     \\ \cline{2-7}
    &1D     &1.03   &1,338      &1,631,465    &26,225,882    &2,873.98     \\ \cline{2-7}
    &2D     &1.01   &  486      &1,632,317    &18,820,634    &1,137.80     \\ \cline{2-7}
    &CRVC   &1.00   &  262      &1,632,541    &34,418,293    &  612.12     \\ \cline{2-7}
    &SC     &1.05   &2,471      &1,630,332    &26,222,846    &2,978.84     \\ \cline{2-7}
    &DC     &1.07   &5,202      &1,627,601    &26,294,017    &3,457.81     \\
    \hline

    \multirow{6}{*}{\parbox[t]{2mm}{\rotatebox[origin=c]{90}{RoadNet-CA}}} 
    &RVC    &1.01   &71          &1,965,135    &10,952,275   &   136.83     \\ \cline{2-7}
    &1D     &1.01   &34          &1,965,172    & 7,484,190   &    83.93     \\ \cline{2-7}
    &2D     &2.85   &0           &1,965,206    &10,414,276   &15,734.31     \\ \cline{2-7}
    &CRVC   &1.02   &63          &1,965,143    & 5,509,769   &   198.77     \\ \cline{2-7}
    &SC     &1.01   &77          &1,965,129    & 7,484,802   &    85.03     \\ \cline{2-7}
    &DC     &1.01   &77          &1,965,129    & 7,484,802   &    85.03     \\
    \hline

    \multirow{6}{*}{\parbox[t]{2mm}{\rotatebox[origin=c]{90}{Orkut}}} 
    &RVC    &1.00   &   48      &3,072,393    &201,579,891   & 1,388.70    \\ \cline{2-7}
    &1D     &1.03   &5,134      &3,067,307    &139,337,672   &22,320.95    \\ \cline{2-7}
    &2D     &1.01   &   13      &3,072,428    & 58,110,415   & 8,115.76    \\ \cline{2-7}
    &CRVC   &1.00   &  206      &3,072,235    &136,670,436   & 1,944.62    \\ \cline{2-7}
    &SC     &1.03   &1,384      &3,071,057    &139,339,199   &22,968.71    \\ \cline{2-7}
    &DC     &1.03   &1,384      &3,071,057    &139,339,199   &22,968.71    \\
    \hline

    \multirow{6}{*}{\parbox[t]{2mm}{\rotatebox[origin=c]{90}{socLiveJournal}}} 
    &RVC    &1.00   &   16      &4,847,555    &91,727,814    &  732.58     \\ \cline{2-7}
    &1D     &1.04   &6,967      &4,840,604    &57,194,270    &6,375.69     \\ \cline{2-7}
    &2D     &1.01   &  180      &4,847,391    &44,408,628    &2,891.27     \\ \cline{2-7}
    &CRVC   &1.00   &  374      &4,847,197    &63,830,074    &1,021.94     \\ \cline{2-7}
    &SC     &1.04   &9,866      &4,837,705    &57,208,429    &5,973.63     \\ \cline{2-7}
    &DC     &1.04   &2,896      &4,844,675    &58,398,861    &7,245.30     \\
    \hline

    \multirow{6}{*}{\parbox[t]{2mm}{\rotatebox[origin=c]{90}{Follow\_jul}}}  
    &RVC    &1.00   &       12  &17,172,130   &77,909,024    &    943.02   \\ \cline{2-7}
    &1D     &8.59   &1,743,865  &15,428,277   &47,218,160    &743,525.59   \\ \cline{2-7}
    &2D     &1.72   &   13,166  &17,158,976   &44,263,566    &201,600.81   \\ \cline{2-7}
    &CRVC   &1.00   &      206  &17,171,936   &70,607,064    &  1,171.01   \\ \cline{2-7}
    &SC     &8.59   &1,471,524  &15,700,618   &47,441,754    &756,719.19   \\ \cline{2-7}
    &DC     &4.26   &  623,712  &16,548,430   &61,962,146    &344,698.67   \\
    \hline

    \multirow{6}{*}{\parbox[t]{2mm}{\rotatebox[origin=c]{90}{Follow\_dec}}} 
    &RVC    &1.00   &       66  &26,339,905  &119,503,790    &    1,134.83 \\ \cline{2-7}
    &1D     &10.0   &2,282,728  &24,057,243  & 65,446,694    &1,333,180.21 \\ \cline{2-7}
    &2D     &2.04   &    9,140  &26,330,831  & 67,832,243    &  377,868.97 \\ \cline{2-7}
    &CRVC   &1.00   &      717  &26,339,254  &107,795,617    &    1,403.23 \\ \cline{2-7}
    &SC     &10.0   &2,161,293  &24,178,678  & 65,517,506    &1,364,952.64 \\ \cline{2-7}
    &DC     &4.86   &  525,799  &25,814,172  & 98,839,823    &  749,972.78 \\
    \hline
\end{tabular}
}
\end{center}
\caption{Partitioning metrics for 256 partitions}
\label{tab:256metrics}
\end{table*}

\end{appendices}
\end{document}